\documentclass[aps,prl,twocolumn,superscriptaddress,showpacs]{revtex4}
\usepackage[]{graphicx}
\usepackage{amssymb,latexsym}

\begin{document}

\title{Longterm Influence of Inertia on the Diffusion of a Brownian Particle}

\author{Giuseppe Pesce}
\email{giuseppe.pesce@fisica.unina.it}
\affiliation{Dipartimento di Fisica Universit\`a degli studi di Napoli, Complesso Universitario Monte S. Angelo, Via Cintia 80126, Napoli, Italy}

\author{Giorgio Volpe}
\affiliation{Institut Langevin, ESPCI ParisTech, CNRS UMR7587, 1 rue Jussieu, 75005 Paris, France}

\author{Giovanni Volpe}
\affiliation{Soft Matter Lab, Department of Physics, Bilkent University, Ankara 06800, Turkey}
\affiliation{UNAM -- National Nanotechnology Research Center, Bilkent University, Ankara 06800, Turkey}

\author{Antonio Sasso}
\affiliation{Dipartimento di Fisica Universit\`a degli studi di Napoli, Complesso Universitario Monte S. Angelo, Via Cintia 80126, Napoli, Italy}
\affiliation{CNR, Istituto Nazionale di Ottica, Sezione di Napoli, Via Campi Flegrei, 34, 80078 Pozzuoli, Italy}

\date{\today}

\begin{abstract}
We demonstrate experimentally that a Brownian particle is subject to inertial effects at long time scales. By using a blinking optical tweezers, we extend the range of previous experiments by several orders of magnitude up to a few seconds. The measured mean square displacement of a freely diffusing Brownian particle in a liquid shows a deviation from the Einstein-Smoluchowsky theory that diverges with time. These results are consistent with a generalized theory that takes into account not only the particle inertia but also the inertia of the fluid surrounding the particle. This can lead to a bias in the estimation of the diffusion coefficient from finite-time measurements. We show that the decay of the relative error is polynomial and not exponential and, therefore, can have significant effects at time scales relevant for experiments.
\end{abstract}


\maketitle

Often a single parameter is key to the description of the motion of a Brownian particle: the particle's diffusion coefficient $D$. Its correct estimation assumes therefore a pivotal importance in many soft matter systems that are characterized thanks to the observation of the motion of a microscopic probe, such as the measurement of nanoscopic forces \cite{RohrbachRSI04,PesceEPL09,VolpePRE07} and of the thermodynamic properties of microscopic systems \cite{ImparatoPRE07,golding2006physical,wang2009anomalous,barkai2012strange}.

In the original picture conceived by Albert Einstein, microscopic particles undergo a never-ending random motion due to collisions with the molecules of the fluid where they are immersed \cite{Einstein1905}. This picture is formalized by the 
original Langevin equation \cite{Langevin1908,Coffey96}:
\begin{equation}\label{eq:inertial}
\underbrace{m \ddot{x}(t)}_{\mathrm{inertia}} = \underbrace{-\gamma \dot{x}(t)}_{\mathrm{friction}} + \underbrace
{\gamma \sqrt{2 D} W(t)}_{\mathrm{diffusion}},
\end{equation}
where $x(t)$ is the particle's position, $m$, $\gamma$ and $D$ are respectively its mass, friction coefficient and diffusion coefficient, and $W(t)$ is a Gaussian white noise  \cite{Oksendal03,VolpeAJP13}. The motion of a microscopic particle is, therefore, governed by two counteracting forces: the friction between the particle and the surrounding viscous fluid modeled by the Stokes drag and the random thermal force modeled by the white noise. Einstein's relation
\begin{equation}\label{eq:einstein}
D =  \frac{k_{\rm B} T}{\gamma},
\end{equation}
connects these two forces to the particle's average kinetic energy per degree of freedom, i.e., $k_{\rm B} T/2$, where $k_{\rm B}$ is the Boltzmann's constant and $T$ is the absolute temperature. The relation in Eq.~(\ref{eq:einstein}) shows that the particle's kinetic energy is limited by the dissipation associated to its collisions with the surrounding fluid molecules \cite{Einstein1905}. This process occurs on the time scale of the \emph{momentum relaxation time} $\tau_{\rm m}=m/\gamma$, which, for small colloidal particles is of the order of a few microseconds \cite{VolpeAJP13}. 

When dealing with particles in the low Reynolds number regime \cite{purcell1977life}, the inertial term in Eq.~(\ref{eq:inertial}) is often neglected. This introduces an error in the estimation of $D$, which, nevertheless, decreases exponentially in $\tau$ with characteristic time $\tau_{\rm m}$. Thanks to this exponential decay, considering that $\tau_{\rm m}$ is in the order of microseconds and typical experiments are performed on timescales of milliseconds and longer, this error can be safely ignored in most experiments, and the motion of the particle is often considered overdamped \cite{UhlenbeckPR30}. The particle's motion, however, is also influenced by the surrounding fluid that has to move in order to refill the space left free by the particle's displacement \cite{HinchJFM75,ClercxPRA92,LukicPRE07,FranoschPRE09,GrebenkovPRE14}. This second inertial effect, known as hydrodynamic memory, develops over the time scale of the fluid momentum relaxation time $\tau_{\rm f}=R^2\rho_{\rm f}/\eta$, where $\rho_{\rm f}$ is the density of the fluid and $\eta$ its viscosity \cite{GrebenkovPRE14}. The effect of the fluid inertia only decays polynomially with $\tau^{-{1\over2}}$ and,  therefore, has a longterm influence on the determination of $D$.

Until now, most attention has been devoted to explore these inertial effects at very short time scales, i.e., $t \approx \tau_{\rm m},\, \tau_{\rm f}$ \cite{LukicPRL05,JeneyPRL08,li2010measurement,HuangNatPhys11,FranoschNat11}, while the polynomial weak decrease of the hydrodynamic memory means that these effects can have influences at time scales that are comparable to those of standard experiments, i.e., up to several seconds. Here, by measuring the diffusive motion of a Brownian particle over several seconds, we experimentally demonstrate that the polynomial weak decay of the correction to the diffusion coefficient due to the presence of hydrodynamic memory produces measurable effects on time scales several orders of magnitude longer than those previously reported.

In order to formalize the above discussion, we consider the mean square displacement (MSD) of a particle. The MSD quantifies how far a particle moves from its initial position and is an experimentally measurable quantity closely related to the particle's diffusion coefficient $D$. In one dimension, the MSD can be calculated as the time-average of the particle's position $\mathrm {MSD}(\tau) =\overline{x(t+\tau) x(t)}$ \cite{barkai2012strange,VolpeAJP13}, where the overbar represents time average, and $D$ can be estimated as
\begin{equation}\label{eq:dlim}
D = \lim_{\tau \rightarrow +\infty} \frac{\mathrm{MSD}(\tau)}{2 \tau} \approx \frac{\mathrm{MSD}(\tau)}{2 \tau},
\end{equation}
where $\tau$ is large. In practice, $\tau$ in Eq.~(\ref{eq:dlim}) cannot be taken to infinity, as it is limited by the experimentally accessible data range. For a massless particle ($m=0$), whose motion obeys Eq.~(\ref{eq:inertial}), the estimation of the diffusion coefficient using Eq.~(\ref{eq:dlim}) is always exact at any $\tau$ and is not affected by experimental limitations in the acquisition of long data series. For a particle with inertia ($m \neq 0$), this estimation leads to an exponentially decreasing error in $\tau$. Nonetheless, this error can be safely neglected in most experiments as it decays on a time scale given by the momentum relaxation time $\tau_{\rm m}$, which, as we have seen, is on the order of microseconds.

\begin{figure}[Hb]
\begin{center}
\includegraphics[width=1\linewidth]{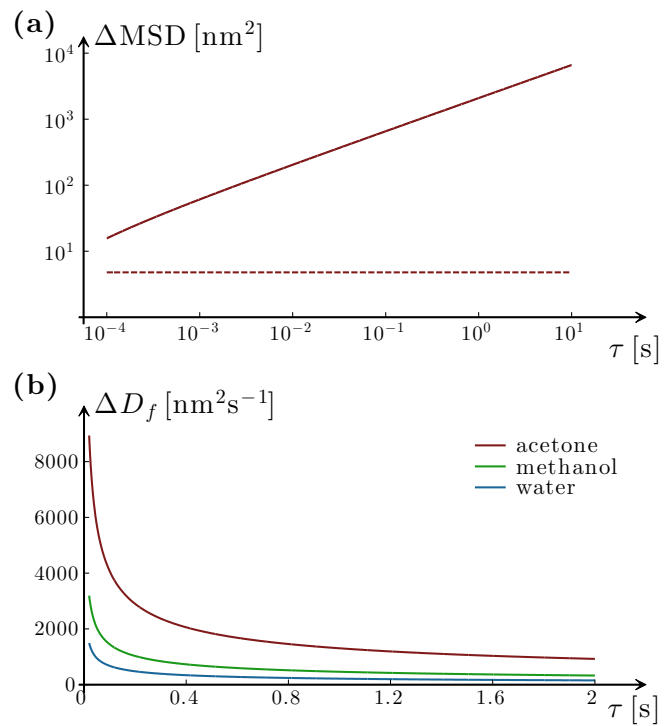}
\caption{(a) Error in the estimation of the ${\rm MSD}(\tau)$ in the case of a melamine particle ($d=8.1\,\mu m$) taking into account  its mass (dashed line) and taking into account also the fluid inertia (solid line). While the former quickly reaches a plateaux, the latter diverges as a function of $\tau$.
(b) Error in the estimation of the diffusion coefficient $D$ for the same particle immersed in the various fluids used in this work taking into account the presence of the fluid inertia [Eq.~(\ref{eq:d_f})], which decays polynomially with $\tau^{-{1\over2}}$. When only the particle inertia is taken into account, the error on the estimation of $D$ decays exponentially with characteristic time $\tau_{\rm m}$ so that the error is negligible on the plotted scales.}
\label{fig:theory}
\end{center}
\end{figure}

However, the MSD for a particle whose motion obeys the Langevin equation corrected to take into account the hydrodynamic  memory effect \cite{HinchJFM75,ClercxPRA92,GrebenkovPRE14} is
\begin{equation}\label{eq:msd_f}
\mathrm{MSD}(\tau) = 2D\tau \bigg\{ 1 - \sqrt{\frac{4 \tau_{\rm f}}{\pi \tau}} - \frac{8\tau_{\rm f}}{9 \tau} + \frac{\tau_{\rm m}}
{\tau}  + \epsilon(\tau) \bigg\},
\end{equation}
where $\epsilon(\tau)$ is a correction term relevant only for very short times, i.e., for $\tau \ll \tau_{\rm f},\, \tau_{\rm m}$, and, thus, can be safely neglected in the following discussion. In this case, as shown by the solid line in Fig.~\ref{fig:theory}(a), the error on the MSD diverges as a function of $\tau$. The error in the estimation of $D$ can be derived using Eq.~(\ref{eq:dlim}) and is
\begin{equation}\label{eq:d_f}
\Delta D_f(\tau) = D-D_f(\tau)= D \left( \sqrt{\frac{4 \tau_{\rm f}}{\pi \tau}} + \frac{8\tau_{\rm f}}{9 \tau} - \frac{\tau_{\rm m}}{\tau} 
\right),
\label{eq.Dmsd}
\end{equation} 
which decays polynomially with $\tau^{-{1\over2}}$. Therefore, this effect can have an influence at time scales comparable to standard experiments, as shown by the solid lines in Fig.~\ref {fig:theory}(b) for the different fluids used in the experiments shown in this article.

\begin{figure}[t]
\begin{center}
\includegraphics[width=1\linewidth]{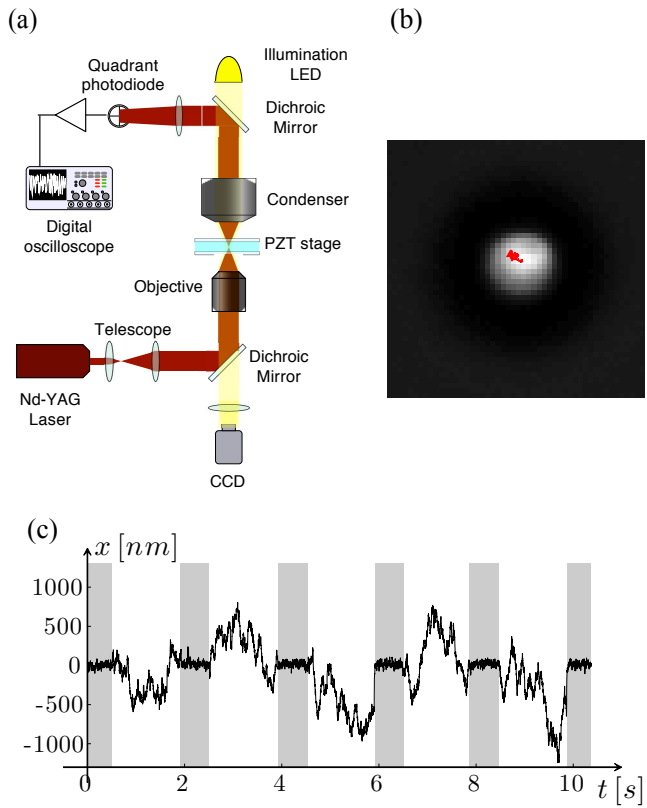}
\caption{(a) The setup consists of a blinking optical tweezers combined with a fast digital video acquisition system. (b) Image and trajectory of a particle freely diffusing after being released from the trap. (c) Example of the $x$-component of the trajectory acquired while the particle is trapped and released repeatedly by the optical tweezers. The grey-shaded areas represent the periods of time when the trapping laser is on.}
\label{fig.setup}
\end{center}
\end{figure}

The experimental setup that we used in order to measure these inertial effects at long time scales is schematically shown in Fig.~\ref{fig.setup}(a). It consists of an optical tweezers built on a high-stability home-made inverted optical microscope equipped with a high-numerical-aperture water-immersion objective lens (Olympus, UPLAPO60XW3, NA=1.20). The optical trap was generated  by a frequency and amplitude stabilized Nd-YAG laser ($\mathrm{\lambda=1.064\,\mu m}$, $\mathrm{500\,mW}$ maximum output power, Innolight Mephisto). The laser beam was expanded to a final diameter of $10\,\mathrm{mm}$ to have a proper overfilling condition. The laser power was precisely regulated rotating a half-wave-plate before a polarizing beam splitter. Finally, the whole setup was enclosed in a thermally and acoustically insulated box to prevent vibrations, air circulation and temperature drifts. A white cold LED was used to illuminate the sample while avoiding heating. The temperature inside the box was continuously monitored: during each set of measurements the temperature fluctuations were within $0.2^{\circ}\mathrm{C}$. To measure the free motion of particles, we employed the blinking optical tweezers technique \cite{CrockerPRL94,PesceOE10}. The laser beam was switched on and off using a home-made electronic shutter driven by an analog waveform generator.  The modulation frequency varied between $0.1\,\mathrm{Hz}$ up to a few Hz depending on the size of the particle used. Videos of the particle motion were acquired using a fast CCD camera with a rate up to 500 fps (see Figs.~\ref{fig.setup}(b) and \ref{fig.setup}(c)). We varied the relevant parameters using three types of particles material ---~polystyrene, melamine, silica and Licristar$^{\circledR}$ (Merck)~--- and three different fluids ---~water, acetone and methanol. Different combinations of these materials allowed us to study the inertial effects under a range of densities and viscosities. The particles were highly diluted to have about few tens of them in the sample cell, i.e., 5$\div$10~particles/microliter. The relevant properties of the particles and fluids used in this work are reported in Table \ref{tab.beads} and \ref{tab.fluids}, respectively. 

\begin{table}[htdp]
\caption{Properties of the particles used in the experiments. Errors on diameters are reported by the manufacturers.}
\begin{center}
\begin{tabular}{ccc}
\hline
Material 		& $d~({\rm \mu m})$  	& $\rho~({\rm g/cm^3})$  \\ 
\hline
Polystyrene 	& 5.0$\pm$0.1 & 1.05 \\ 
Melamine 		& 8.1$\pm$0.1 	& 1.51 \\ 
Licristar$^{\circledR}$ 		& 3.10$\pm$0.05 	& 1.41 \\ 
Silica 		&4.8$\pm$0.2 	& 1.80 \\ 
\hline
\end{tabular}
\end{center}
\label{tab.beads}
\end{table}

\begin{table}[htdp]
\caption{Properties of the fluids used in the experiments. }
\begin{center}
\begin{tabular}{ccc}
\hline
Fluid		& $\rho~({\rm g/cm^3})$	& $\eta~({\rm mPa\, s})$ \\
\hline
Water	& 1.00 &  0.89 \\
Methanol 	& 0.79 & 0.60 \\
Acetone 	& 0.79 & 0.30  \\ 
\hline
\end{tabular}
\end{center}
\label{tab.fluids}
\end{table}

\begin{figure}[b]
\begin{center}
\includegraphics[width=1\linewidth]{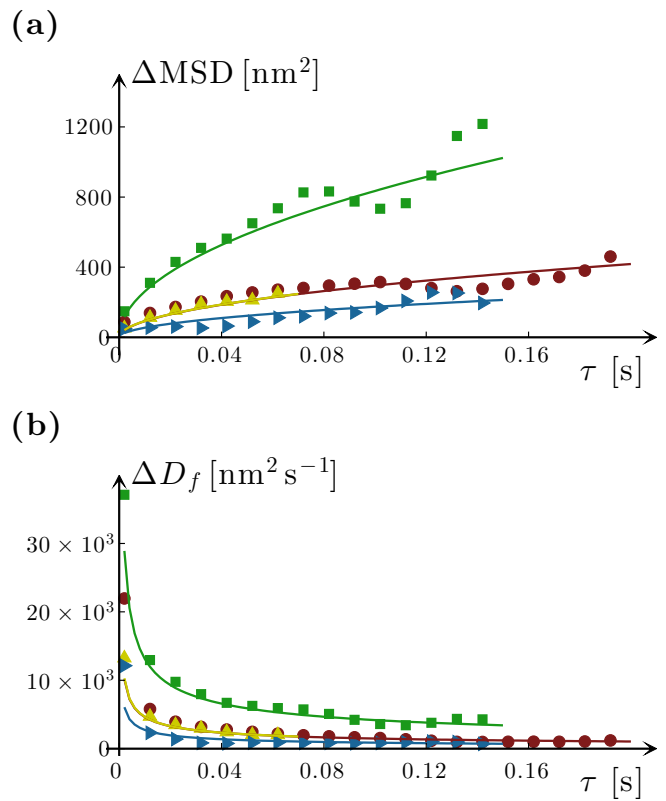}
\caption{Errors in the determination of the ${\rm MSD}(\tau)$ (a) and of the diffusion coefficient $D$ (b) for different combinations of particles and fluids: (squares) 3.1 $\mu m$ Licristar bead in acetone, (circles) 5.0  $\mu m$ polystyrene bead in methanol, (up triangles) 4.8 $\mu m$ silica bead in methanol, (right triangles) 3.1 $\mu m$ Licristar bead in water. The experimental errors are similar to those reported in Fig.~\ref{fig.longdev} and are not shown for clarity.}
\label{fig.shortdev}
\end{center}
\end{figure}

To demonstrate the polynomial increase of the error on the measurements of the MSD, we estimated it from a large number of recorded trajectories of a freely diffusing particle. Fig.~\ref{fig.shortdev}(a) shows the difference of the MSD calculated using the Stokes-Einstein relation with that obtained from experimental values for different particle sizes in different fluids and Fig.~\ref{fig.shortdev}(b) shows the corresponding errors in the estimation of the value of $D$. The main contribution to the error is due to the viscosity; in fact, going from water to methanol to acetone the error increases. This is related to the fact that in low viscosity fluids particles diffuse much faster than in high viscosity fluids and  the effect of the fluid inertia is larger. It is interesting to note that the effect is clearly visible also in water where most of experiments relying on diffusion are done. The particles' density, instead, is not a crucial parameter. In fact, the MSD deviation of the $5\,{\rm \mu m}$ polystyrene and silica particles in the same fluid (methanol) lie on the same theoretical curve, even though the density of silica is roughly twice that of polystyrene.

\begin{figure}[t]
\begin{center}
\includegraphics[width=1\linewidth]{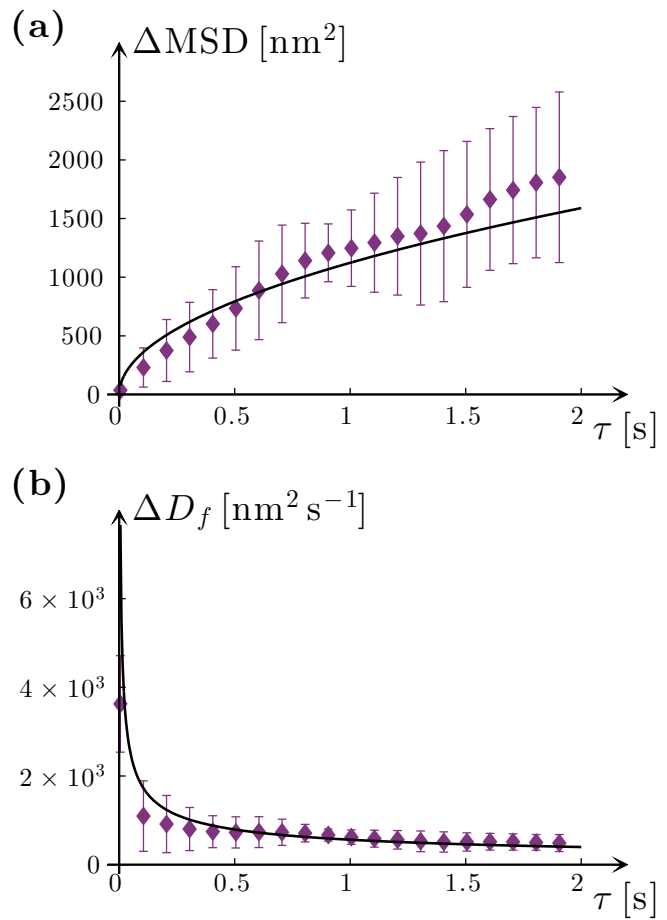}
\caption{The polynomial behaviour of the error in the estimation of (a) the ${\rm MSD}(\tau)$ and (b) of the diffusion coefficient $D$ for a $8.1\,{\rm \mu m}$  melamine particle up to few seconds.}
\label{fig.longdev}
\end{center}
\end{figure}

The results presented in Fig.~\ref{fig.shortdev} are limited to $150\,{\rm ms}$ since for heavy particles the frequency of the blinking optical tweezers need to be set high enough in order to limit the fall of the particle due to effective gravity to a range where the blinking optical tweezers is still able to re-trap it. In order to overcome this limitation and to measure the deviation for times up to a few seconds, we used melamine particles (diameter $8.1\,{\rm \mu m}$) in acetone. Due to their weight these particles immediately reach the bottom coverslip, where they perform a quasi-two-dimensional random motion. The particles are prevented from sticking because of the presence of electrostatic repulsive forces. Finally, we used the blinking optical tweezers technique to place the particle at a given initial position and digital video microscopy to record its trajectory. The blinking frequency was set to $0.125\,{\rm Hz}$. For each particle, we recorded about 5000 trajectories at $200\,{\rm fps}$ and calculated the corresponding MSD for each trajectory and then the average. We then averaged the results obtained with five different particles. The resulting deviation between the experimentally measured MSD and the one corresponding to the free diffusion of a massless particle is presented in Fig.~\ref{fig.longdev}. We obtained a very good agreement between the experimental data and the expected values, which clearly shows that the polynomial increase of the error is an appreciable effect up to few seconds.

In conclusion, we have demonstrated experimentally that a Brownian particle is subject to inertial effects at long time scales, extending the range of previous experiments by several orders of magnitude up to a few seconds. The measured MSD of a freely diffusing Brownian particle in a liquid shows a deviation from the Einstein-Smoluchowsky theory that diverges with time. These results are consistent with a generalized theory that takes into account the displacement of the fluid surrounding the particle. This can lead to a bias in the estimation of the diffusion from finite-time measurements, as the decay of the relative error is polynomial and not exponential it can have significant effects at time scales relevant for experiments.

\begin{acknowledgments}
This work was partially supported by the MPNS COST Action 1205 ``Advances in Optofluidics: Integration of Optical Control and Photonics with Microfluidics." Giovanni Volpe was partially supported by Marie Curie Career Integration Grant (MC-CIG) PCIG11 GA-2012-321726.
\end{acknowledgments}


\end{document}